\documentclass[12pt]{article}

\usepackage{amssymb, amsmath, amsfonts, natbib, graphicx}
\usepackage[margin=1in]{geometry}
\usepackage{bm, float, rotating, setspace}

\usepackage{color, url}

\newfloat{algorithm}{htbp}{loa}
\floatname{algorithm}{Algorithm}

\DeclareMathOperator{\E}{E}

\newtheorem{result}{Result}

\title{Diagnostic tools for approximate Bayesian computation using the coverage property}

\author{D. Prangle\footnote{Mathematics and Statistics Department, Lancaster University, Lancaster, U.K.}\;, M. G. B. Blum\footnote{Universit\'e Joseph Fourier, Centre National de la Recherche Scientifique, Laboratoire TIMC-IMAG UMR 5525, Grenoble, F-38041, France}\;,  G. Popovic\footnote{School of Mathematics and Statistics, University of New South Wales, Sydney, Australia}\; and S. A. Sisson$^\ddagger$\footnote{Email: {\tt Scott.Sisson@unsw.edu.au}}}

\date{}

\begin{document}

\maketitle

\begin{abstract}

Approximate Bayesian computation (ABC) is an approach for sampling from an approximate posterior distribution in the presence of a computationally intractable likelihood function. A common implementation is based on simulating model, parameter and dataset triples, $(m,\theta,y)$, from the prior, and then accepting as samples from the approximate posterior, those pairs $(m,\theta)$ for which $y$, or a summary of $y$, is ``close'' to the observed data.
Closeness is typically determined though a distance measure and a kernel scale parameter, $\epsilon$. 
Appropriate choice of $\epsilon$ is important to producing a good quality approximation.
This paper proposes diagnostic tools for the choice of $\epsilon$ based on assessing the coverage property, which asserts that credible intervals have the correct coverage levels.  We provide theoretical results on coverage for both model and parameter inference, and adapt these into diagnostics for the ABC context. We re-analyse a study on human demographic history to determine whether the adopted posterior approximation was appropriate. {\tt R} code implementing the proposed methodology is freely available in the package {\tt abc}.

\end{abstract}

\section{Introduction}
\label{sec:intro}

For a given model, $m$, Bayesian inference for unknown parameters $\theta$, and given observed data $y_{\text{obs}}$, updates prior beliefs $\pi(\theta|m)$ through the likelihood function $\pi(y_{\text{obs}} | \theta,m)$. This produces the posterior distribution $\pi(\theta | y_{\text{obs}},m) =  \pi(y_{\text{obs}} | \theta,m)\pi(\theta|m) / \pi(y_{\text{obs}}|m)$, where $\pi(y_\text{obs}|m) = \int  \pi(y_{\text{obs}} | \theta,m)\pi(\theta|m) d \theta$ is  the integrated likelihood for model $m$.  Similarly, Bayesian inference for a discrete set of models $m \in \{1,2,\ldots,M\}$ updates a prior mass function $p(m)$ to posterior weights $p(m | y_{\text{obs}}) \propto \pi(y_{\text{obs}} | m)p(m)$.
Here the joint posterior of parameter and model is given by
$\pi(\theta,m|y_{\text{obs}})\propto \pi(y_{\text{obs}}|\theta,m)\pi(\theta|m)p(m)$.
Increased usage of Bayesian inference in recent decades has been built on powerful algorithms, such as Markov chain Monte Carlo, which make use of repeated evaluation of the likelihood function(s).

Approximate Bayesian computation (ABC) refers to a family of algorithms which perform an approximate Bayesian inference when numerical evaluation of the likelihood is not feasible, but where it is possible to simulate from the model(s) $y\sim\pi(\cdot|\theta,m)$.  
ABC has become a popular tool for the analysis of complicated models in a wide range of challenging applications.  See e.g.  \cite{Beaumont:2010, Bertorelle:2010, Csillery:2010, Marin:2011} and \cite{sisson+f11} for overviews of methods and applications.

A common, importance sampling-based implementation of ABC, expressed for the multi-model setting, is given by the following:
\begin{algorithm}[ht]
\begin{footnotesize}
\vspace{1mm}
\rule{\textwidth}{0.1mm}
\underline{ABC importance sampling}

\begin{enumerate}
\item 
For $i=1,\ldots,N$:\\
Sample a model and parameter from the prior $(\theta_i,m_i)\sim \pi(\theta|m)p(m)$.\\
Simulate data from model $m_i$ as $y_i \sim \pi(y | \theta_i,m_i)$.
\item \label{step2} Weight each sample $(y_i,\theta_i,m_i)$ by $w_i\propto K_\epsilon(\|y_i-y_{\text{obs}}\|)$.
\end{enumerate}

\rule{\textwidth}{0.1mm}
\caption{\small An ABC importance sampling algorithm, based on a single large sample of size $N$, incorporating model choice and parameter inference. \label{alg:ABC}}
\end{footnotesize}
\end{algorithm}

Here $K_\epsilon(u)=K(u/\epsilon)/\epsilon$ is a standard smoothing kernel with scale parameter $\epsilon>0$, and $\|\cdot\|$ is a distance measure e.g. Euclidean. For this article, for simplicity and w.l.o.g. we consider $K_\epsilon(u)$ to be the uniform kernel $U(-\epsilon,\epsilon)$, so that step \ref{step2} above corresponds to selecting (i.e. with non-zero weights) those samples $(y_i,\theta_i,m_i)$ for which $\|y_i-y_{\text{obs}}\|\leq\epsilon$.
Note that when $p(m)$ is small, the above algorithm may have large Monte Carlo error in estimating $p(m | y_{\text{obs}})$.  This is often avoided by using a uniform prior mass function in place of $p(m)$ as a computational device, and then reweighting each model appropriately (see e.g. \cite{Grelaud:2009}).

The output of the above algorithm is a sample of parameter vectors $(\theta_i,m_i)$ from an approximation to the posterior
\begin{equation}
\label{eqn:abcapprox}
\pi_{\text{ABC}}(\theta,m | y_{\text{obs}}) \propto \pi(\theta|m)p(m) \int \pi(y | \theta,m) K_\epsilon(\|y - y_{\text{obs}}\|) \, dy,
\end{equation}
where it can be seen that  $\pi_{\text{ABC}}(\theta,m | y_{\text{obs}})\approx\pi(\theta,m | y_{\text{obs}})$ following standard conditional density estimation arguments.
There are two sources of approximation error in the sample representation of (\ref{eqn:abcapprox}) from Algorithm \ref{alg:ABC}, both of which are influenced by $\epsilon$.  The first is the discrepancy between $\pi_{\text{ABC}}(\theta,m | y_{\text{obs}})$ and $\pi(\theta,m | y_{\text{obs}})$.  These are equal in the limit $\epsilon \to 0$,  however the approximation deteriorates as $\epsilon$ is increased.  The extreme result is $\lim_{\epsilon \to \infty} \pi_{\text{ABC}}(\theta, m | y_{\text{obs}}) = \pi(\theta|m)p(m)$.  That is, all proposed samples $(\theta,m)$ are accepted, so that ABC targets the prior distribution and all information from the data is lost.  
Secondly, $\pi_{\text{ABC}}(\theta,m | y_{\text{obs}})$ is approximated by a finite sample whose size reduces as $\epsilon \to 0$ (in other ABC algorithms this corresponds to extremely low acceptance rates).  Indeed the sample size typically reduces to zero for continuous data.  In effect, $\epsilon$ controls a form of the usual bias-variance trade-off \citep{Blum:2010}.

Many approaches have been proposed to reduce the approximation error.  One is to replace $\|y - y_{\text{obs}}\|$ by $\|s - s_{\text{obs}}\|$ where $s=S(y)$ is a vector of summary statistics.
Low dimensional but informative summary statistics can greatly improve inferential accuracy, even at the price of potential information loss \citep{Blum:2012}.  Using summary statistics introduces another level of approximation, as then ABC  approximates $\pi(\theta,m|s_{\text{obs}})$ rather than $\pi(\theta,m|y_{\text{obs}})$.
A second approach is to post-process the sample from $\pi_{\text{ABC}}(\theta,m|y_{\text{obs}})$ with $\epsilon>0$ so that is approximately transformed to a sample with $\epsilon=0$.  Termed regression-adjustment, for within-model parameter inference this takes the form of linear or non-linear regression-based transformations of $\theta_i$ \citep{Beaumont:2002,Blum/Francois:2010, Blum:2012}. For the adjustment of model probabilities, post-processing can be performed by multinomial regression \citep{Beaumont:2008}.

\subsection{Diagnostics for ABC}

This paper addresses two related open questions about ABC in practice.  Firstly, is it possible to validate the ABC approximation of the posterior, $\pi_{\text{ABC}}(\theta,m|y_{\text{obs}})\approx \pi(\theta,m|y_{\text{obs}})$ (or $\pi_{\text{ABC}}(\theta,m|s_{\text{obs}})\approx \pi(\theta,m|s_{\text{obs}})$ when using summary statistics), as accurate?  Secondly, how should $\epsilon$ be chosen?  Typically $\epsilon$ is  commonly chosen in an ad-hoc manner, although several authors \citep{Bortot:2007, Ratmann:2009, Blum:2010b,faisai+fh13} have suggested approaches where $\epsilon$ is estimated as part of an extended model.

In this paper we approach the question of the accuracy of the ABC posterior approximation by examining whether the \emph{coverage property} holds (described below).  By numerically evaluating adherence to the coverage property through diagnostic statistics, we are able to determine the likely accuracy of the ABC approximation for a range of $\epsilon$ values.
In favourable circumstances this allows the user to choose as large a value of $\epsilon$ as possible (for computational efficiency) for which coverage approximately holds.  Alternatively, the coverage diagnostics may reveal that large approximation error remains for any choice of $\epsilon$.

Our approach is based on  credible intervals, a standard Bayesian method to give interval parameter estimates.  
Consider the case of within-model parameter inference for a fixed model $m_0$.
An $\alpha$\% credible interval for a univariate parameter $\theta$ is an interval $I$ with the property that $\Pr(\theta \in I | y_{\text{obs}}) = \alpha/100$.  Suppose that  credible intervals are constructed from ABC output for data simulated from a known parameter value, $\theta_0$.  Roughly speaking, the coverage property asserts that these intervals have the claimed probability of containing $\theta_0$.  To fully define the property we must be specific about the distribution of $\theta_0$.  This is discussed later, where we give theoretical results supporting a particular choice for our purpose.  We also describe how this definition can be extended to model inference problems.

An equivalent condition to the coverage property was provided by \cite{Cook:2006}.  Namely, the $p$-values of $\theta_0\leq\theta$ (or $\theta_0>\theta$) within the posterior estimates must have a $U(0,1)$ distribution. Accordingly, the coverage property may be tested for a particular value of $\epsilon$, by repeatedly performing ABC for many choices of $\theta_0$ and associated pseudo-observed data $y_0 \sim \pi(y|\theta_0, m_0)$, computing $p$-values, and then applying standard tests for uniformity.  We provide a similar equivalent condition and a test of the coverage property for model probabilities, in addition to providing a computationally efficient process to compute the test statistics, for both within- and between-models.

There is a sizeable literature related to the coverage property.  Bayesian work includes determining the correctness of complex Bayesian simulation algorithms \citep{Cook:2006}, 
the post-processing of within-model ABC output \citep{menendez+fgs12}
and the validation of ABC analyses in the single model setting \citep{Wegmann:2009, Wegmann:2010,Aeschbacher:2012}.
A recent overview of work from a frequentist perspective is provided by \cite{Gneiting:2007}.  However, this work has the somewhat different aim of determining consistency between statistical predictions and a sequence of observed outcomes (e.g.~weather forecasts and meteorological data).  
 Despite the difference in aims, the primary ideas behind our statistical tests of coverage go back to the frequentist literature: \cite{Dawid:1984} for continuous parameters and \cite{Seillier:1993} for model choice.
The approach we develop in this article is similar to the ABC papers mentioned above.  Our contribution here, is to explain and justify the theoretical basis and methodology of coverage in more detail, to improve this methodology where needed, and to extend these ideas to the hitherto unconsidered realm of model inference for ABC.

The remainder of the paper is structured as follows:  
Section \ref{sec:coverage} defines the coverage property for both parameter and model inference, and gives some theoretical results. The methodological details of the resulting diagnostics are described in Section \ref{sec:method}, including an algorithm and discussion of diagnostic statistics and tests.  Section \ref{sec:example} presents a simulated example to illustrate the methods and justify some implementation choices, followed by a re-analysis of a study into human demographic history to determine whether a reliable posterior approximation was obtained.
Finally, Section \ref{sec:discussion} concludes with a discussion.

\section{Coverage} 
\label{sec:coverage}

We investigate whether the ABC approximation $\pi_{\text{ABC}}(\theta,m|y_{\text{obs}})$ (or $\pi_{\text{ABC}}(\theta,m|s_{\text{obs}})$) is a good representation of the posterior $\pi(\theta,m|y_{\text{obs}})$ (or $\pi(\theta,m|s_{\text{obs}})$) by testing the \emph{coverage property}.  For inference on a continuous scalar parameter, $\theta$, an informal definition is that a given credible interval based on $\theta|y_0$, where $y_0\sim\pi(y|\theta_0,m_0)$ for fixed $m_0$, should contain the true parameter, $\theta_0$, the appropriate proportion of times.
This Section presents a more precise definition, a discussion of the property's consequences, and results on how it can be tested.  We also describe a version of the property suitable for a model choice setting. We notationally work with $y$ rather than $s$ throughout this Section.

\subsection{Parameter inference}
\label{sec:sec}

We define the coverage property for the case of a continuous scalar parameter $\theta$ for a fixed model $m_0$, where for the remainder of Section \ref{sec:sec} we drop all notational dependence on $m_0$.  For multivariate parameter vectors, our method will examine each parameter separately.

The informal definition above is based on analysing data $y_0$ simulated from known parameter values $\theta_0$.  To formalise the property we introduce a distribution for these, $H(\theta_0,y_0)$, with densities associated with $H$ denoted by $h$.
A natural choice, used by \cite{Cook:2006}, \cite{Wegmann:2009, Wegmann:2010} and \cite{Aeschbacher:2012}, is $h(\theta_0, y_0) = \pi(y_0 | \theta_0) \pi(\theta_0)$; draw the parameters from the prior, and the data from the model of interest conditional on this.  We present an argument in favour of an alternative choice for the ABC setting below.

Let $g(\theta|y)$ be a density approximating the posterior given data $y$.  From this, credible intervals of any level can be constructed.  Suppose $C(y, \alpha)$ is a credible interval of level $\alpha$\%.  
We say that $g$ satisfies coverage with respect to $H$ if the coverage level of such an interval is $\alpha$ when analysing data generated from $H$  (i.e.~$\Pr(\theta_0 \in C(y_0, \alpha)) = \alpha$),  for any choice of $\alpha$ and $C$.
More formally, we have:

\paragraph{Definition of coverage property:}

Let $g(\theta|y)$ be a density approximating the univariate posterior $\pi(\theta|y)$, and $G_y(\theta)$ be the corresponding distribution function.  Consider a function $B(\alpha) \subseteq [0,1]$ defined for $\alpha \in [0,1]$ such that the resulting set has Lebesgue measure $\alpha$.  Let $C(y, \alpha) = G_y^{-1}[B(\alpha)]$ and $H(\theta_0,y_0)$ be the distribution function for $(\theta_0,y_0)$.  We say $g$ satisfies the coverage property with respect to distribution $H(\theta_0, y_0)$ if for every function $B$ and every $\alpha \in [0,1]$, $\Pr(\theta_0 \in C(y_0, \alpha)) = \alpha$.
\vspace{4mm}

There are two requirements for the coverage property to be a useful criterion to determine how well $g(\theta|y)$ approximates the posterior $\pi(\theta|y)$.  These requirements will determine some characteristics of $H(\theta_0,y_0)$. Firstly, it should hold when $g(\theta|y)=\pi(\theta|y)$.  

\begin{result} \label{res:postcov_pi}
The posterior, $\pi(\theta | y)$, satisfies coverage with respect to any distribution $H(\theta_0, y_0)$ with conditional density $h(\theta_0|y_0)=\pi(\theta_0 | y_0)$.
\hfill {\rm Proof in Appendix.}
\end{result}

The second requirement is that the coverage property should avoid false positives: it should not hold when $g(\theta|y)\neq\pi(\theta|y)$.  However, coverage can hold when $g(\theta|y)=\pi(\theta)$ equals the prior distribution,  when $\theta_0\sim\pi(\theta)$ is also drawn from the prior.  

\begin{result} \label{res:priorcov_pi}
The prior, $\pi(\theta)$, satisfies coverage with respect to any distribution $H(\theta_0, y_0)$ with marginal density $h(\theta_0)=\pi(\theta_0)$.
\hfill {\rm Proof in Appendix.}
\end{result}

The above results demonstrate that the choice $h(\theta_0, y_0) = \pi(\theta_0, y_0)$ (where  $\pi(\theta_0, y_0)=\pi(\theta_0|y_0)\pi(y_0)=\pi(y_0|\theta_0)\pi(\theta_0)$) leads to the coverage property holding for both the prior and posterior distributions.  
The false positive of the prior is particularly unwelcome in the ABC context, as it corresponds exactly to the ABC approximation $\pi_{\text{ABC}}(\theta|y)$ for $\epsilon \to \infty$. (The prior also coincides with the ABC posterior  approximation when $s=S(y)$ has no information for $\theta$ under the model, so that $\pi_{\text{ABC}}(\theta|s)=\pi(\theta|s)=\pi(\theta)$, although this is due to Result \ref{res:postcov_pi} rather than Result \ref{res:priorcov_pi}. See Section \ref{sec:discussion} for more discussion.)

To avoid this we propose the alternative choice of $h(\theta_0, y_0) \propto \pi(\theta_0, y_0) \mathbb{I}[y_0 \in A]$.  That is, the distribution $\pi(\theta_0, y_0)$ truncated to require that the data lie within some subset $A$.  This preserves $h(\theta_0|y_0)=\pi(\theta_0|y_0)$,
so coverage holds for the posterior, but it typically alters $\pi(\theta_0)$ so that coverage not longer holds for the prior (i.e. $h(\theta_0)\neq\pi(\theta_0)$).  We examine some convenient choices of $A$ in Section \ref{sec:method}.  
In this manner, we are aiming to evaluate coverage for datasets similar to $y_{\text{obs}}$, rather than the much stronger context of coverage holding for all datasets.
Note that the above results do not prove that the posterior $\pi(\theta|y)$ is the only distribution to satisfy coverage with respect to our choice of $H$. However, we are unaware of any other such distributions that are likely to arise in the ABC context.

An equivalent condition to the coverage property, which is easier to test,  is the following:

\begin{result} \label{res:equivcov_pi}
Let $H$ be the distribution function of $(\theta_0, y_0)$.  Define $p_0 = G_{y_0}(\theta_0)$, where $G_y(\theta)$ is the distribution function of $\theta$ under $g(\theta|y)$.  Coverage holds with respect to $H$ iff 
\begin{equation}
\label{eqn:ptest1}
p_0 \sim U(0,1).
\end{equation}
\hfill {\rm Proof in Appendix.}
\end{result}
A similar result was given by \cite{Cook:2006}, who proved that under coverage (with respect to  the distribution function  $h(\theta_0, y_0) = \pi(y_0 | \theta_0) \pi(\theta_0)$) the empirical distribution of $p_0$  converges to $U(0,1)$.

\subsection{Model inference}

As for parameter inference, the definition of the coverage property for model inference requires us to specify the distribution of the known parameter values $m_0$ and $y_0$ through $H(m_0, y_0)$, which can be considered a marginal distribution of $H(m_0,\theta_0,y_0)$. Note that for model inference, $H$, its derivatives ($h$), and the mass function $g(m|y)$  approximating the posterior $p(m|y)$ are discrete functions.

Formalising the intuitive notion of coverage for the case of model choice faces the difficulty of interpreting the idea of a credible interval for a discrete parameter.  We firstly illustrate our definition with an example, and then formalise it below.  Suppose that given data $y_0$ simulated from model $m_0 \in \{1,2,3\}$, estimated posterior probabilities are $0.7, 0.2$ and $0.1$.  This could be viewed as defining three credible intervals; a $70\%$ credible interval that $m=1$ etc.  We would like to investigate coverage in the following sense: given a $70\%$ interval for $m=1$ produced by some $(m_0, y_0)$ pair, there is a $70\%$ probability of it containing $m_0$.  A technical difficulty is that the probability of a pair producing a $70\%$ credible interval is typically zero.  This difficulty can be avoided by requiring the following condition to hold for every $a<b$ e.g.~$a=0.69$ and $b=0.71$:  Consider all $y_0$ such that the estimated probability of $m=1$ is between $a$ and $b$.  Conditioning on this, the probability of $m_0=1$ also lies between $a$ and $b$.

\paragraph{Definition of coverage property:}

Let $g(m|y)$ be a mass function approximating the posterior and $G_y(m)$ the corresponding distribution function.  Given $I=[a,b] \subseteq [0,1]$, define $A(m,I) = \{ y | g(m| y) \in I \}$.  We say $g$ satisfies the coverage property with respect to distribution $H(m_0, y_0)$ if, for all $i\in\{1,2,\ldots,M\}$ and $I$, either
\begin{align}
\Pr(y_0 \in A(i,I)) &= 0, \qquad \text{or} \label{eq:Apos} \\
\Pr(m_0=i | y_0 \in A(i,I)) &\in I. & \label{eq:mc_cov}
\end{align}

Similar arguments to the parameter inference case show that the posterior satisfies coverage when $h(m_0 | y_0) = \pi(m_0 | y_0)$, but the prior satisfies coverage when $h(m_0) = p(m_0)$.  

\begin{result} \label{res:postcov_mc}
The posterior, $p(m | y)$, satisfies coverage with respect to any distribution $H(m_0, y_0)$ with conditional mass function $h(m_0|y_0)=p(m_0 | y_0)$. 
\hfill 
{\rm Proof in Appendix.}
\end{result}

\begin{result} \label{res:priorcov_mc}
The prior, $p(m)$, satisfies coverage with respect to any distribution $H(m_0, y_0)$ with marginal mass function $h(m_0)=p(m_0)$.
\hfill  {\rm Proof in Appendix.}
\end{result}

This means that the natural choice of $h(m_0,y_0)=p(m_0,y_0)$ (where $p(m_0,y_0)=p(y_0|m_0)p_0(m_0) = p(m_0|y_0)p(y_0)$)
is not suitable.  As before, our proposed solution is to truncate this distribution on $(m_0, y_0)$, so that $h(m_0,y_0)=p(m_0,y_0)\mathbb{I}[y_0\in A]$.

As before, the above definition of coverage for model inference is not directly testable. Below we give an equivalent (under weak technical conditions) form which is.

\begin{result} \label{res:equivcov_mc}
Define $z_0^{(i)} = g(i| y_0)$.  Assume that for all $i\in\{1,2,\ldots,M\}$, the measure $\mathcal{Z}^{(i)}$ on $z_0^{(i)}$ induced by the distribution $H(m_0,y_0)$ is not a singular continuous distribution with respect to Lebesgue measure.  Coverage then holds with respect to $H(m_0,y_0)$ if and only if, for all $i$
\begin{equation} \label{eqn:ptest2}
\Pr(m_0=i | z_0^{(i)}=w) = w
\end{equation}
holds for almost all $w$ with respect to $\mathcal{Z}^{(i)}$.
\hfill {\rm Proof in Appendix.}
\end{result}

\section{Method} 
\label{sec:method}

In this Section we discuss how to construct diagnostics based on the coverage property. In principle, this is simply a matter of repeatedly constructing an ABC posterior approximation, $\pi_{\text{ABC}}(m,\theta|y_0)$, for known values of $(m_0,\theta_0,y_0)\sim H(m_0,\theta_0,y_0)$, computing $p$-values and estimated model probabilities, and then testing whether the conditions \eqref{eqn:ptest1} and \eqref{eqn:ptest2} hold. This can be repeated for many $\epsilon$ values until a suitable choice is found.
However, simulating datasets for a single ABC analysis is typically computationally expensive. As such, we reuse the same simulations for each ABC analysis, along the lines of Algorithm \ref{alg:ABC}, as is common for ABC diagnostics (e.g. \cite{Blum:2012}). Simulations in Section \ref{sec:example} indicate this makes little difference to the results.

In the following, we first present the full algorithm, including how to generate $(m_0, \theta_0, y_0)\sim H(m_0,\theta_0,y_0)$. We then describe several test statistics and diagnostic plots to allow an assessment of whether the conditions \eqref{eqn:ptest1} and \eqref{eqn:ptest2} hold.
As in the previous Section, the details are presented in terms of $y$ rather than $s$ for notational simplicity.
{\tt R} code to implement these methods has been made available as part of the {\tt abc} package \citep{Csillery:2012}.

\subsection{Algorithm}
 \label{sec:algorithm}

\begin{algorithm}[hpt]
\begin{footnotesize}
\rule{\textwidth}{0.1mm}
\underline{ABC coverage diagnostics}
\begin{enumerate}
\item Determine integers $N>0$, $c>0$ and candidate values of $\epsilon$: $\epsilon_1 > \epsilon_2 > \ldots \epsilon_q\geq 0$. 
\item Simulate a set $\mathbb{U}=\{(m_i, \theta_i, y_i) | i=1,\ldots, N\}$ of independent realisations of $(m,\theta,y)$ from $\pi(y|\theta,m)\pi(\theta|m)p(m)$.
\item Select $\mathbb{V} \subseteq \mathbb{U}$ containing the $c$ realisations that minimise $\|y_i-y_{\text{obs}}\|$.
\item For each $(m_0, \theta_0, y_0) \in \mathbb{V}$ and for $j=1,\ldots,q$:
\begin{enumerate}
  \item \label{step3a} Let $\mathbb{W} = \mathbb{U} \setminus (m_0, \theta_0, y_0)$.
  \item Find the subset of $\mathbb{W}$ such that $\|y_i-y_{\text{obs}}\| \leq \epsilon_j$.
  \item (Optional) Perform regression-adjustment post-processing.
  \item Record $p$-values and estimated model probabilities.
\end{enumerate}
 \item Construct plots of diagnostic statistics versus $\epsilon$.
\end{enumerate}
\rule{\textwidth}{0.1mm}
\caption{\small An algorithm to diagnose coverage for ABC as a function of kernel scale parameter $\epsilon>0$. In the case of a single model, modify $(m,\theta,y)\rightarrow(\theta,y)$ and $\pi(y|\theta,m)\pi(\theta|m)p(m)\rightarrow\pi(y|\theta)\pi(\theta)$ in the obvious way.} \label{alg:mc}
\end{footnotesize}
\end{algorithm}

The algorithm for diagnosing coverage of the ABC approximation $\pi_{ABC}(m,\theta|y_{\text{obs}})$ (or $\pi_{\text{ABC}}(\theta|y_{\text{obs}})$) is presented in Algorithm \ref{alg:mc}.
The set $\mathbb{V}$ is a sample of size $c$ from the distribution $h(m_0,\theta_0,y_0)=\pi(y_0|\theta_0,m_0)\pi(\theta_0|m_0)p(m_0)\mathbb{I}[y_0\in A]$, where $A=\{y:\|y-y_{\text{obs}}\|\leq\delta\}$ for some $\delta$ determined by $c$. Each element of $\mathbb{V}$ is taken as the known values $(m_0,\theta_0,y_0)$ in turn, and the ABC posterior approximation estimated for a range of kernel scale parameters, $\epsilon$.

Increasing $c$, the number of known values of $(m_0,\theta_i,y_0)$, will improve the power of the tests of coverage. However the tradeoff is a greater computing cost, and that $A$ becomes less concentrated around $y_{\text{obs}}$, so the risk of the prior satisfying the coverage property increases.  The final choice is left to the user. However, we note that $c$ can be increased (or decreased) based on preliminary findings.  We investigate various values of $c$ by simulation in Section \ref{sec:example}, and based on this suggest $c=200$ as a default.

\subsection{$P$-values and model probabilities} %
\label{sec:summarising}

For scalar $\theta$, we require a $p$-value estimate of (\ref{eqn:ptest1}) under Result \ref{res:equivcov_pi}, based on a posterior sample $\theta^{(1)}, \theta^{(2)},\ldots,\theta^{(n)}$.
Here, we use
$p = (1 + \sum_{j=1}^n \mathbb{I}(\theta^{(j)} < \theta_0))/(2+n)$,
which is equivalent to the posterior mean for the binomial probability that $\theta_j < \theta_0$ under a uniform prior.  
This choice eliminates the occurrence of extreme values ($p=0$ or $1$), which can overly influence some test statistics.
For multivariate $\theta$, we record a $p$-value estimate for each parameter.

Given a posterior sample of model indicators $m^{(1)},m^{(2)},\ldots,m^{(n)}$, a straightforward estimate of the posterior probability of model $i$ is the proportion which equal this: $g(m=i|y_0)=\sum_{j=1}^n \mathbb{I}(m^{(j)}=i)/n$. Alternatively, regression-adjusted post-processing produces estimated posterior model probabilities directly \citep{Beaumont:2008}.

Removing an element of $\mathbb{U}$ in step \ref{step3a} of Algorithm \ref{alg:mc} can slightly bias the estimated ABC model probabilities.  For example, let $d = \sum_{i=1}^n \mathbb{I}[m_i=1]$.  For $\epsilon = \infty$, $g(m=1|y_0) = (d-1)/(n-1)$ for $m_0=1$ and $d/(n-1)$ otherwise.  This dependence of $g(\cdot|y_0)$ on $m_0$ causes unwanted behaviour in some diagnostics.  To mitigate this, we reweight the model probability estimates to use the empirical prior model weights from $\mathbb{U}$ rather than $\mathbb{W}$.  That is, given estimated model probabilities $g(m_i|y_0)$ we adjust these to $\tilde{g}(m=i|y_0) \propto g(m=i|y_0) h_i(\mathbb{U}) / h_i(\mathbb{W})$, where $h_i(\cdot)$ gives the proportion of realisations from model $i$ in the supplied set.  No similar correction of parameter estimates was found to be necessary.

\subsection{Diagnostic statistics} %
\label{sec:diagstats}

For each parameter and value of $\epsilon$ we will have $c$ replicated $p$-values, $p_1,p_2,\ldots,p_c$. We treat these as independent, although there may be mild dependence induced by Algorithm \ref{alg:mc}. Under the coverage property these will be distributed as $U(0,1)$ (Result \ref{res:equivcov_pi}). There are a number of tests for uniformity.
\cite{Cook:2006} used the diagnostic statistic
\begin{equation}
\label{eq:CGR}
X^2 = \sum_{i=1}^c (\Phi^{-1} (p_i))^2,
\end{equation}
where $\Phi$ is the standard normal distribution function.  When the $p_i$ values are independent $U(0,1)$ draws, $X^2 \sim \chi^2_c$, which allows the calculation of a $p$-value for coverage (we report the $p$-value for a two-tailed test).  We note that this statistic is unchanged if some $p_i$ values are replaced with $1-p_i$; it does not test for symmetry of the distribution around $0.5$.  This can cause problems in practice.  An example based on real data is the top left histogram of Figure S5
which displays $p_i$ values that are clearly not uniform but receive a $p$-value of 0.75 from this diagnostic statistic.

An alternative used by \cite{Wegmann:2009, Wegmann:2010} is the Kolmogorov-Smirnoff test statistic
\begin{equation} \label{eq:KS}
Y = \sup_x |F_c(x) - F(x)|,
\end{equation}
where $F_c(x)$ is the empirical distribution function of $p_1, p_2, \ldots, p_c$ and $F(x)$ is the $U(0,1)$ distribution function.  The distribution of $Y$ 
if $p_i\sim U(0,1)$
is known in exact and asymptotic forms \citep{Durbin:1973} and can be used to calculate a $p$-value for coverage (using a one-tailed test).  Our $p_i$ values are not drawn from continuous distributions, but rather discrete distributions based on the number of posterior samples.  However asymptotic $p$-values based on the continuous distribution can still be calculated and will be of the correct order of magnitude, which suffices for their purpose as a rough diagnostic guide.  If more accurate $p$-values are required, Monte Carlo estimation is possible but more time consuming.

For model inference diagnostics, we focus on the binary case of model $m=i$ and model $m\neq i$, for each $i\in\{1,\ldots,M\}$.
For each $\epsilon$, Algorithm \ref{alg:mc} is run on a sequence of $y_0$ values, $y_{0,1}, y_{0,2}, \ldots, y_{0,c}$, generated from $m_0$ values, $m_{0,1}, m_{0,2}, \ldots, m_{0,c}$, to produce $z$ values, $z_1, z_2, \ldots, z_c$, where $z_j$ is the estimated probability of model $i$: $\Pr_{\text{ABC}}(m=i | y_{0,j})$.
Define $q_j = \mathbb{I}(m_{0,j}=i)$.  Following Result \ref{res:equivcov_mc}, we wish to test the coverage hypothesis that $q_j \sim \text{Bernoulli}(z_j)$, where all $q_j$ values are assumed independent, as before.

A simple diagnostic statistic is the proportion of times model $i$ occurs,
\[
U = c^{-1} \sum_{j=1}^c q_j.
\]
A central limit theorem holds for the distribution of $U$ under the null hypothesis of coverage, conditional on the $z_j$ values.  However, this can be a poor approximation when some $z_j$ values are close to 0 or 1.  Instead, we construct the null distribution by Monte Carlo methods to estimate the $p$-value for coverage (using a two-tailed test).  To improve the stability of the resulting $p$-values, we use the same random seed across different $\epsilon$ values.

A drawback of $U$ is that highly unlikely $q_j$ values, such as $q_j=1$ when $z_j=10^{-6}$ provide strong evidence against coverage, but do not contribute more to $U$.  As an alternative, we can consider the log-likelihood,
\begin{equation} \label{eq:loglik}
V = \sum_{j=1}^c \left[ q_j \log z_j + (1-q_j) \log(1-z_j) \right],
\end{equation}
with $p$-values for coverage (using a two-tailed test) again calculated by Monte Carlo simulation.
A drawback of this statistic is that $V$ is constant regardless of the $q_i$ values if $z_j \equiv 0.5$, 
and so coverage cannot be rejected.
Also note a similar statistic is to use the log-likelihood of $c$ independent discrete random variables, $W = \sum_{j=1}^c \log \Pr_\text{ABC}(m=m_{0,j} | y_{0,j})$.  This tests coverage of all models.

A problem with these statistics is that they can be insensitive to departures from coverage which vary in nature with $q_j$.  
It is difficult to define a statistic which is flexible enough to detect such problems for all possible $(q_j, z_j)$ sequences.  \cite{Seillier:1993} present a portmanteau statistic combining tests on a partition chosen for a specific dataset, but in our experience such a statistic is hard to adapt to work generally.  As such, we advise that checking diagnostic plots is particularly important.  Should these show poor performance of general purpose diagnostic statistics, they may motivate a better choice specific to the problem of interest.

\subsection{Diagnostic plots}
\label{sec:diagplots}

For parameter inference, many
standard diagnostic plots can be used to assess uniformity of $p_1, p_2, \ldots, p_c$, such as histograms and  probability plots.

For model inference, we present a diagnostic plot, an example of which is shown in Figure \ref{fig:BIA_mc_ci}.  
Based on an equally-spaced partition of $[0,1]$ into subintervals, $S$,
we estimate $\Pr(q_j=1 | z_j \in S)$ for each subinterval, by Bayesian inference for a binomial rate using a uniform prior.  The diagnostic plot displays each posterior mean and 95\% credible interval.  Under coverage, each credible interval should include some of the associated $z_j$ interval with high probability.  The plot illustrates whether the coverage property appears to hold for each interval individually.  This approach is similar to the \cite{Seillier:1993} portmanteau statistic described above, but without the need to combine the results into a single statistic.  Indeed, they also propose a ``coverage plot,'' similarly plotting point estimates for several partitions.

\section{Analyses}
\label{sec:example}

\subsection{Simulated example} 

We now examine how our coverage diagnostics perform in a simple simulated example. We consider that 100 data points are drawn independently from either a $N(0,1)$  or a $gk(0,1,g,0)$ model, which are equally likely a priori. (For details on the $g$-and-$k$ distribution see e.g.~\citealt{Drovandi:2011}.)
Inference can be split into binary model choice, and inference for the unknown parameter, $g>0$, which has a $U(0,4)$ prior.  Observed data, $y_{\text{obs}}$, was drawn from the $g$-and-$k$ model with $g=0.2$.  We base our ABC analysis on the median and upper and lower quartiles as summary statistics, so we are interested in determining how well the ABC approximation $\pi_{\text{ABC}}(m,\theta|s_{\text{obs}})$ represents the posterior $\pi(m,\theta|s)$. For the analyses in this paper, we use weighted Euclidean distance 
$\|a-b\|=[ \sum_j (a_j - b_j)^2/v_j^2 ]^{1/2}$, where $v_j^2$ is the prior predictive variance of the $j$-th summary statistic, estimated from the set $\mathbb{W}$ in Algorithm  \ref{alg:mc}.

\begin{sidewaysfigure}[p] \begin{center}
\includegraphics{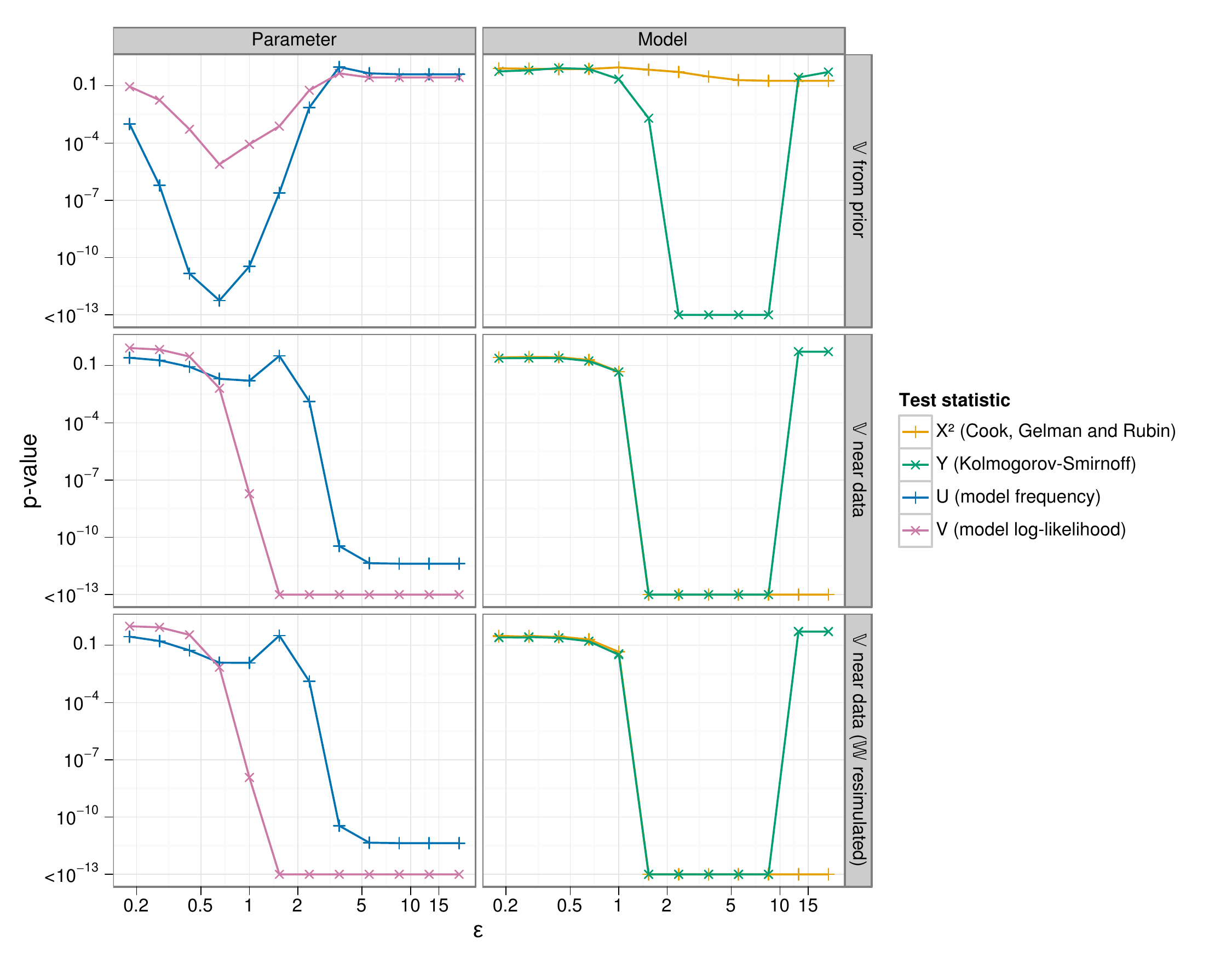}
  \caption{\footnotesize{Plots of $\epsilon$ against coverage $p$-values for the $N(0,1)$ / $g$-and-$k$ example.  Left panels correspond to parameter inference, and right panels for model inference.  Top panels indicate when $\mathbb{V}$ consists of $c=200$ samples drawn from the prior; middle panels when $\mathbb{V}$ is constructed from the $c=200$ samples in which $s$ is closest to $s_{\text{obs}}$.  Top and middle panels use Algorithm \ref{alg:mc} which re-uses $(m,\theta,s)$ samples throughout the analyses. Bottom panels simulate new $(m,\theta,s)$ samples for each ABC analysis.  
   }
}
  \label{fig:gknorm_stats}
\end{center}
\end{sidewaysfigure}

For analysis, we construct the set $\mathbb{U}$ from $2 \times 10^6$ simulated $(m,\theta,y)$ triples, half from each model. 
Figure \ref{fig:gknorm_stats} shows coverage diagnostic $p$-values from the statistics $U$, $V$, $X^2$ and $Y$ as a function of $\epsilon$, and for parameter (left panels) and model (right panels) inference respectively.  For parameter inference, we only present results from the $g$-and-$k$ model. The top panels show diagnostics when the set of known values, $\mathbb{V}=\{(m_0,\theta_0,s_0)\}$, is a random sample of size $c=200$ from $\mathbb{U}$ (i.e. the prior), and the middle and bottom panels when $\mathbb{V}$ consists of the $c=200$ samples with $s$ closest to $s_{\text{obs}}$.

The top panels in Figure \ref{fig:gknorm_stats} support Results \ref{res:priorcov_pi} and \ref{res:priorcov_mc}, in that when $(m_0,\theta_0,s_0)\sim\pi(s|\theta,m)\pi(\theta|m)p(m)$ are drawn from the prior, then coverage holds both for the prior (i.e.~large $\epsilon$ values) and the posterior (i.e.~small $\epsilon$ values). When $(m_0,\theta_0,s_0)\sim\pi(s|\theta,m)\pi(\theta|m)p(m)\mathbb{I}(\|s_0-s_{\text{obs}}\|\leq\delta)$ are drawn from the truncated prior (Figure \ref{fig:gknorm_stats}, middle panels), then coverage does not hold for the prior. Note that the statistic $U$ does not detect any deviation from coverage here, as discussed in Section \ref{sec:diagstats}. Figure \ref{fig:gknorm_stats} also illustrates our earlier point, particularly in the case of parameter inference, that requiring coverage to hold for the prior is more demanding than requiring coverage to hold for the truncated prior. This is evidenced by the upturn in $p$-values only occurring at lower $\epsilon$ values when $\mathbb{V}$ is drawn from the prior, compared to the relatively larger values of $\epsilon$ when $\mathbb{V}$ is drawn from the truncated prior.

For comparison, the procedure with $\mathbb{V}$ drawn from the truncated prior was repeated by resimulating $\mathbb{W}$ for each ABC analysis, thereby removing any effects of reusing $(m,\theta,s)$ samples, albeit at far greater computational expense.  Figure \ref{fig:gknorm_stats} (bottom panels) shows that the results are nearly identical to those obtained using Algorithm \ref{alg:mc}.
Further, we repeated our analysis with $c=100$ and $c=500$, and obtained qualitatively similar results (see Figures S1 and S2 in the Supplementary Information), suggesting that the choice of $c$ is not crucial to drawing the correct inferences.

\begin{figure}[ht] \begin{center}
\includegraphics{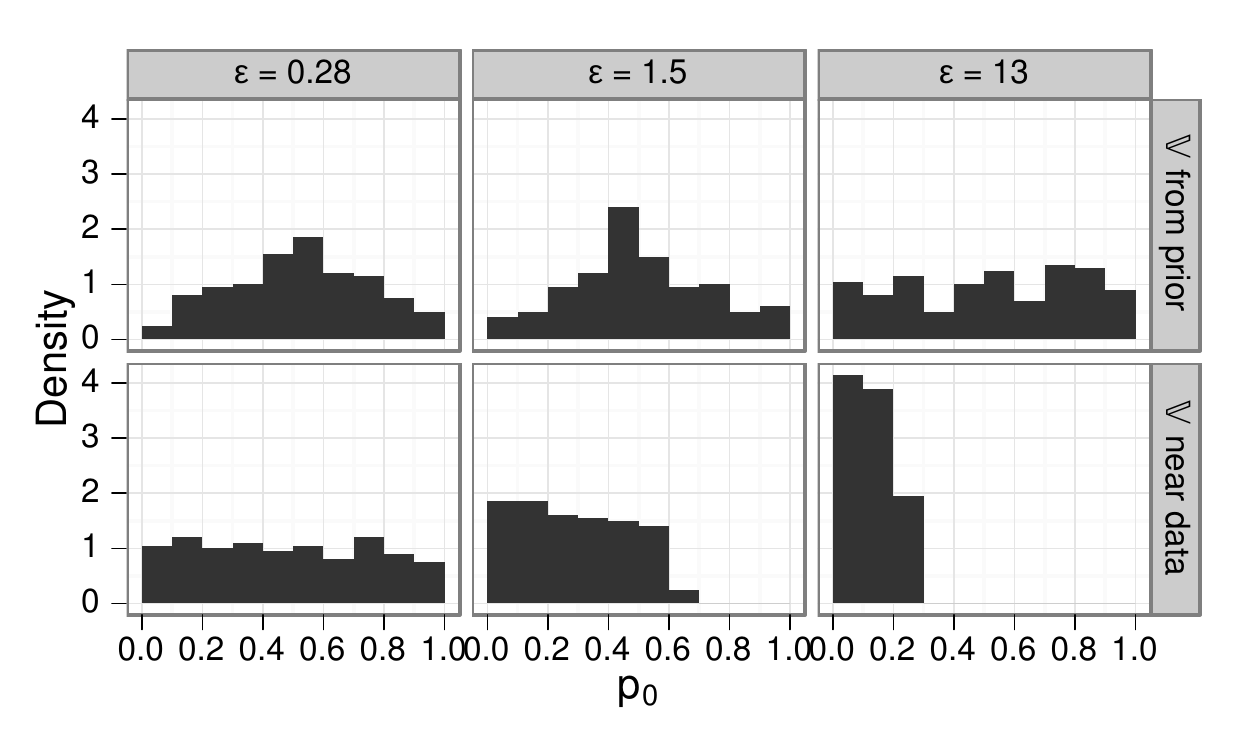}
  \caption{\footnotesize{Histograms of the $c=200$ $p_0$ values for the parameter $g$ in the $N(0,1)$ / $g$-and-$k$ example, for $\epsilon=0.28, 1.5, 13$.  In the top panels $\mathbb{V}$ is drawn from the prior; in the bottom panels $\mathbb{V}$ is drawn from the truncated prior.  Columns indicate different $\epsilon$ values.}
  }
  \label{fig:gknorm_hist}
\end{center}
\end{figure}

\begin{figure}[ht] \begin{center}
\includegraphics{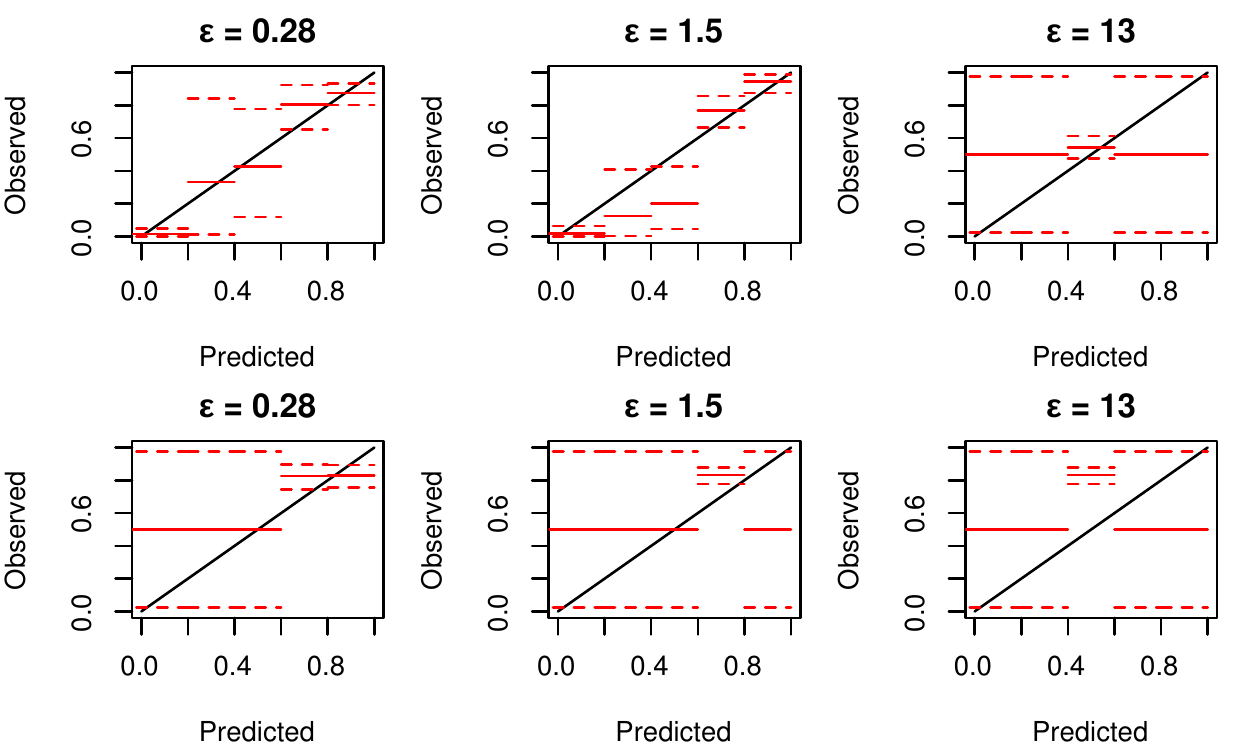}
  \caption{\footnotesize{Model inference diagnostics in the $N(0,1)$ / $g$-and-$k$ example, for $\epsilon=0.28, 1.5, 13$.  In the top panels $\mathbb{V}$ is drawn from the prior; in the bottom panels $\mathbb{V}$ is drawn from the truncated prior.  Columns indicate different $\epsilon$ values.
  Each panel shows the observed and predicted (under coverage) model probabilities for the $N(0,1)$ model, including a 95\% credible interval for the predictions.
  }}
  \label{fig:gknorm_mc}
\end{center}
\end{figure}

Figures \ref{fig:gknorm_hist} and \ref{fig:gknorm_mc} show diagnostic plots to investigate coverage in more detail for $\epsilon=0.28, 1.5, 13$.  These again demonstrate  that coverage approximately holds for large $\epsilon$ when $\mathbb{V}$ is drawn from the prior, but not from the truncated prior. 
They also provide insight into disagreements between the statistics within some panels in Figure \ref{fig:gknorm_stats}.
For parameter inference, in the top left panel of Figure  \ref{fig:gknorm_stats}, there is less than clear agreement about whether coverage roughly holds for the smallest $\epsilon$ values.  The top left panel of Figure \ref{fig:gknorm_hist} confirms that the $p$-value histogram has a non-uniform shape in this case.  
For model inference, the top right panel of Figure \ref{fig:gknorm_stats} suggests no deviation from coverage for any $\epsilon$ for the $U$ statistic.  However, the top centre panel of Figure \ref{fig:gknorm_mc} illustrates that this is not correct.

Our interpretation of these results is that $\epsilon \leq 0.28$ is sufficient to achieve approximate coverage for both parameter inference and model selection in the case where $\mathbb{V}$ is drawn from the truncated prior. Coverage does not hold (for small $\epsilon$) when $\mathbb{V}$ is drawn from the prior, which represents a stricter condition. However, the former case relates more to the dataset and analysis of interest.

\subsection{Application in human demographic history}
 \label{sec:application}

\cite{Sjodin:2012} detail an ABC analysis of genetic data to choose between three demographic models of human history: null, bottleneck and fragmentation models. Each model contains 9 unknown parameters: $b$, $d$, $n$, $N_0$, $N_A$, $N_B$, $T_b$,   $T_g$ and $T_{\text{dur}}$.
Their analysis used $10^5$ simulations from each model, and they accepted the $0.5\%$ of simulations minimising $\|s-s_{\text{obs}}\|$, corresponding to $\epsilon = 1.36$.  \cite{Sjodin:2012} then used regression post-processing for parameter and model inference. 
We use the same experimental setup to evaluate the implemented value of $\epsilon$, and also to determine whether regression-adjusted post-processing improved the results.  

\begin{figure}[!p] \begin{center}
  \includegraphics{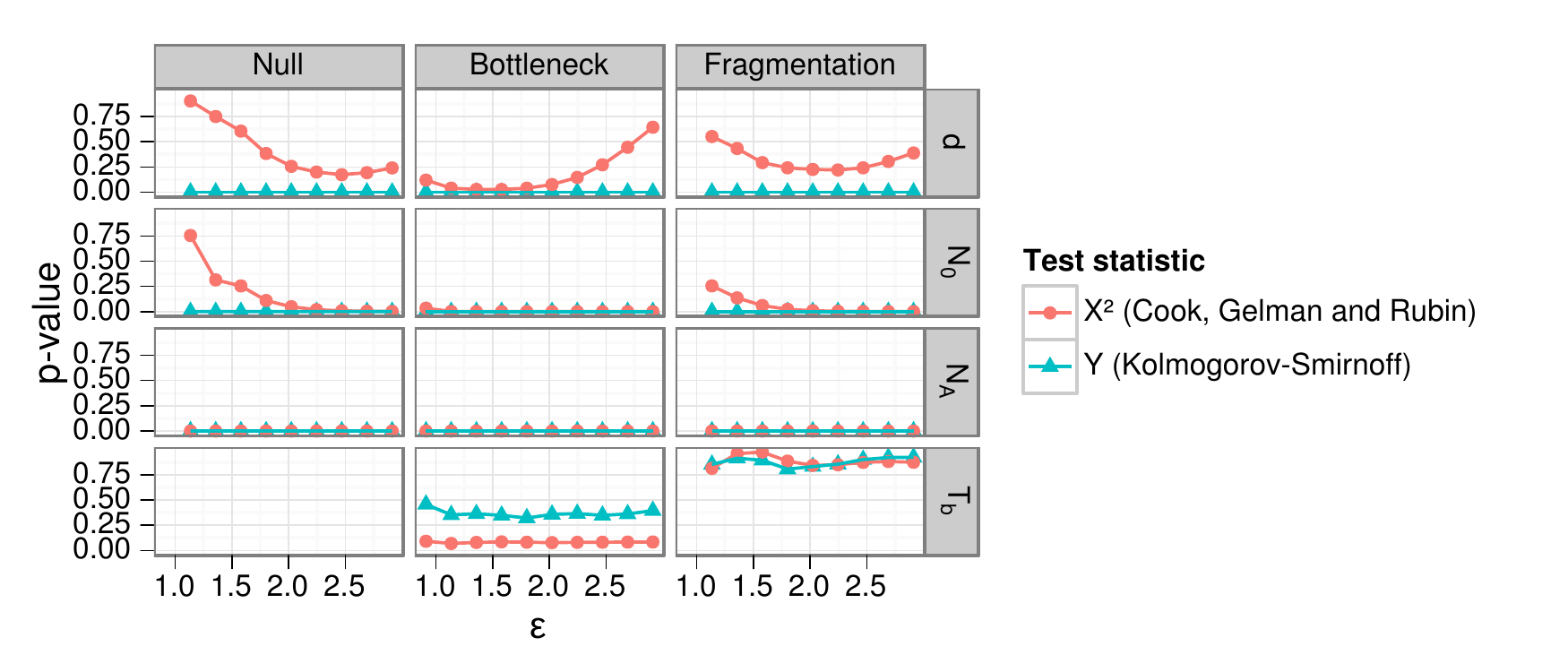}
  \caption{\footnotesize{Plots of $\epsilon$ against coverage $p$-values for the parameters $d$, $N_0$, $N_A$ and $T_b$ in the human demographic history analysis. Regression-adjusted post-processing is not implemented. Rows correspond to individual parameters; columns correspond to the three models.  
 }}
  \label{fig:BIA_pi_nocorr}
\end{center}
\end{figure}

\begin{figure}[ht] \begin{center}
\includegraphics{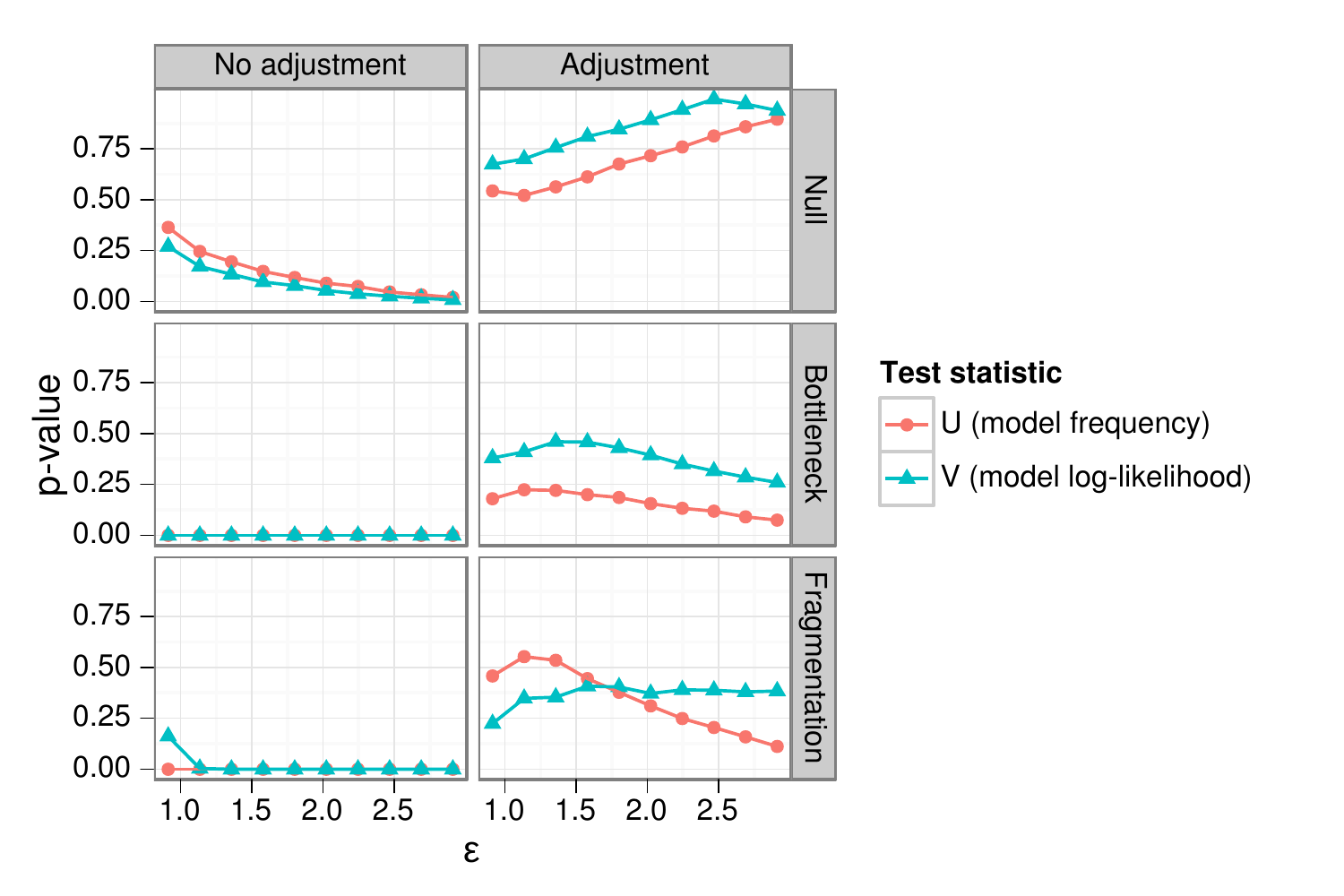}
  \caption{\footnotesize{Plots of $\epsilon$ against coverage $p$-values in the human demographic history analysis. Rows correspond to the three models; columns correspond to the implementation of regression-adjustment post-processing.
   }}
  \label{fig:BIA_mc}
\end{center}
\end{figure}

\begin{figure}[ht] \begin{center}
  \includegraphics{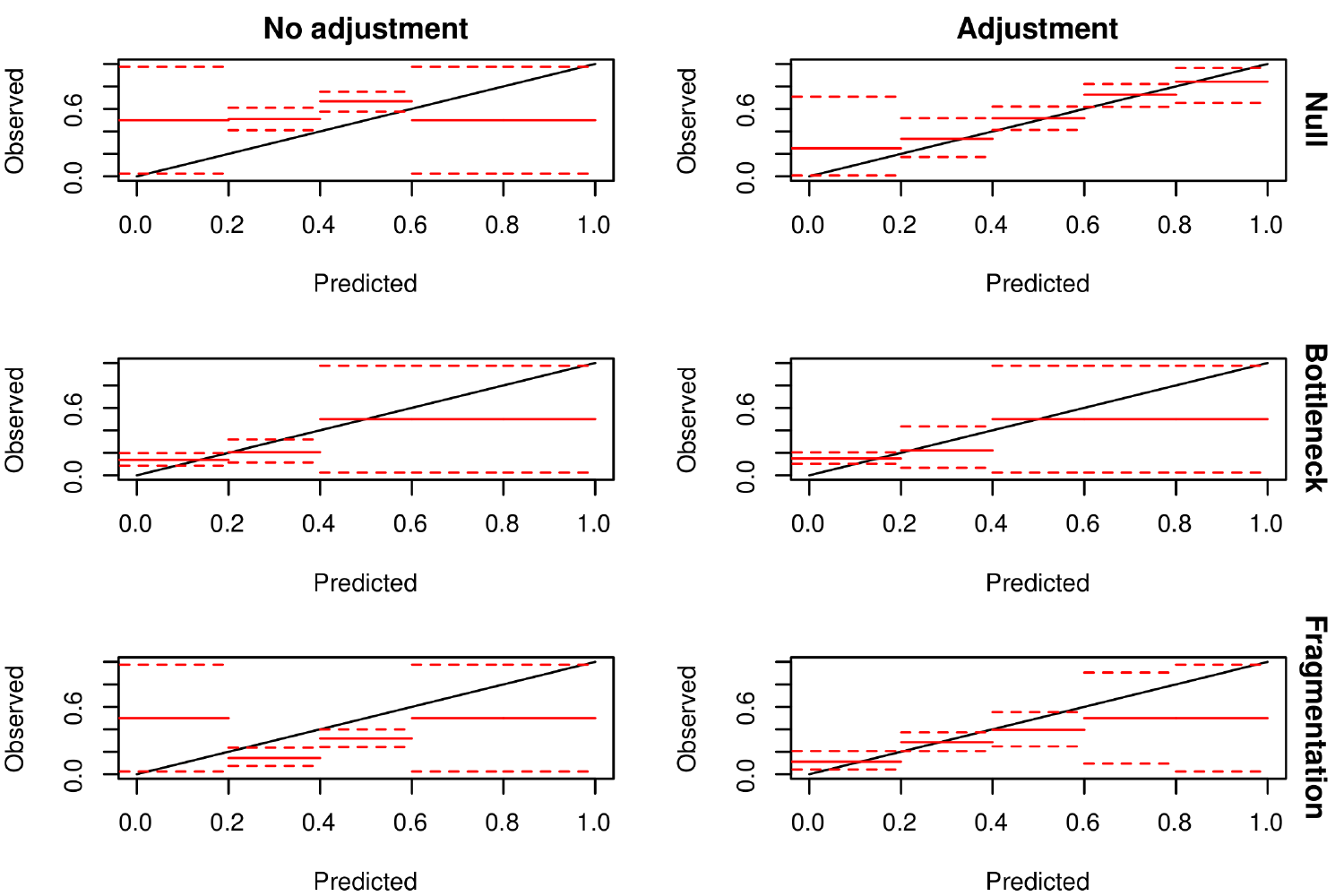}
  \caption{\footnotesize{ Model inference diagnostics in the human demographic history analysis. Rows represent (top) the null, (middle) bottleneck and (bottom) fragmentation models; columns correspond to the implementation of regression-adjustment post-processing.
   Each panel shows the observed and predicted (under coverage) model probabilities for each model, including a 95\% credible interval for the predictions. 
   }}
  \label{fig:BIA_mc_ci}
\end{center}
\end{figure}

Figure \ref{fig:BIA_pi_nocorr} shows parameter inference diagnostics for  $d$, $N_0$, $N_A$ and $T_b$, without regression post-processing (see Supporting Information Figure S3 for the same plot for the remaining parameters).   
There is occasional significant disagreement between the statistics. However, overall it is apparent that coverage is not attained for any $\epsilon$, apart from perhaps $T_{\text{dur}}$.  $P$-value histograms for $\epsilon=1.36$ (Figure S5) confirm that in most cases there is clear deviation from coverage.  For model choice, both statistics agree that coverage is not attained (Figure \ref{fig:BIA_mc}; left panels) and diagnostic plots for $\epsilon=1.36$ confirm this (Figure \ref{fig:BIA_mc_ci}; left panels).

\begin{figure}[!p] \begin{center}
  \includegraphics{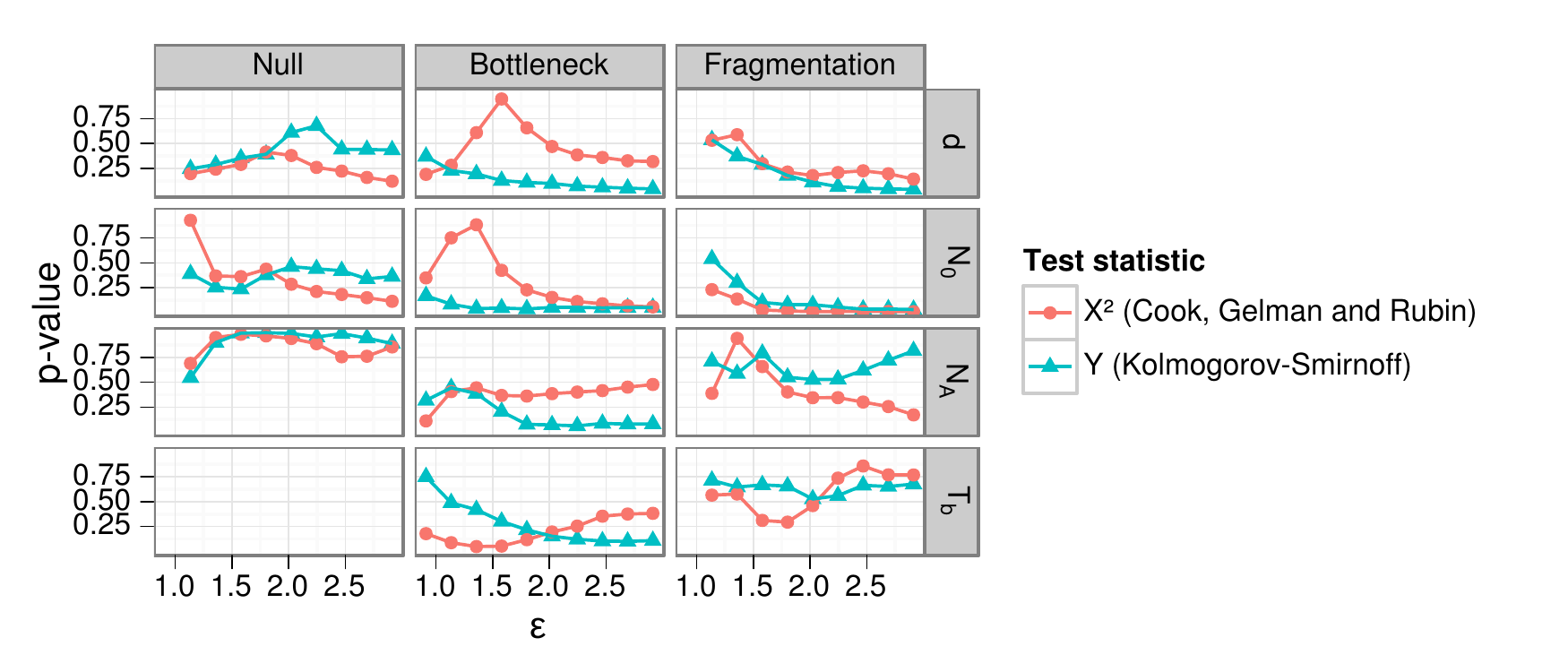}
  \caption{\footnotesize{Plots of $\epsilon$ against coverage $p$-values for the parameters $d$, $N_0$, $N_A$ and $T_b$ in the human demographic history analysis. Regression-adjusted post-processing has been implemented. Rows correspond to individual parameters; columns correspond to the three models.  
  }}
  \label{fig:BIA_pi_corr}
\end{center}
\end{figure}

Regression post-processing was performed by conditional heteroskedastic, local-linear regression \citep{Blum/Francois:2010} for parameter inference, and multinomial logistic regression \citep{Beaumont:2008} for model inference.
The regression post-processing greatly improves the results.  The model inference statistics (Figure \ref{fig:BIA_mc}; right panels) now suggest that coverage holds for any choice of $\epsilon$ investigated.  The same is true of many parameters, although for others, coverage appears to hold only for smaller $\epsilon$ (Figure \ref{fig:BIA_pi_corr}, Figure S4).  Diagnostic plots for parameter and model inference (Figure S6 and \ref{fig:BIA_mc_ci}; right panels) produced for $\epsilon = 1.36$ suggest that coverage is approximately achieved,  except for some small concerns remaining for some parameters (e.g. $N_B$ and $T_{\text{dur}}$). 

On the whole, and with only a few minor caveats, our investigation largely validates the choice of $\epsilon$, and the use of regression post-processing by \cite{Sjodin:2012}.

\section{Discussion}
 \label{sec:discussion}

We have presented a method for validating whether an ABC analysis contains significant approximation error based on assessment of the coverage property.  The method can be used to determine the kernel scale parameter, $\epsilon$, via simple dagnostic plots.
We have used this method in a re-analysis of human demographic data \citep{Sjodin:2012},  validating the choice of $\epsilon$ and the use of regression-adjustment  post-processing in that study.

Our methodology draws on several previous approaches.  In particular, \cite{Wegmann:2009, Wegmann:2010} use a similar scheme for validating ABC parameter inference, and suggest using the $Y$ diagnostic statistic \eqref{eq:KS}. Also, \cite{Cook:2006} employ a similar idea for Bayesian software testing using the $X^2$ statistic \eqref{eq:CGR}.  Our contribution here is to provide: results on the choice of $(m_0,\theta_0,y_0)$ samples to use, a description of a general purpose methodology (and {\tt R} code to implement it), and evidence that many diagnostic statistics are not fully trustworthy and should be supplemented with diagnostic plots.  We also extend the coverage property definition and the validation methodology to cover model inference, incorporating ideas from \cite{Seillier:1993}.

Our approach only aims to determine whether $\pi_{\text{ABC}}(m,\theta|y_{\text{obs}})$ is a good approximation of $\pi(m,\theta|y_{\text{obs}})$, or whether $\pi_{\text{ABC}}(m,\theta|s_{\text{obs}})$ is a good approximation of $\pi(m,\theta|s_{\text{obs}})$. In order to use the coverage property to assess the approximation impact of summary statistics, in addition to that of $\epsilon$, it would be necessary to choose the set $\mathbb{V}$ to consist of those $(m,\theta,y)$ that minimise $\|y-y_{\text{obs}}\|$, and then perform the rest of the analysis based on using $(m,\theta,S(y))$ rather than $(m,\theta,y)$. Further investigation would be required to see if this is a practical approach.

One word of caution: when a summary statistic is not informative for a parameter, so that $\pi_{\text{ABC}}(\theta|s)=\pi(\theta|s)=\pi(\theta)$, then our diagnostics will support a good posterior approximation for any value of $\epsilon$. This is, of course, the correct result, however it should not be misconstrued as any support of information content in $s$ for $\theta$.
Additionally, the above diagnostics are evaluated for each parameter separately within a multivariate parameter $\theta$. Hence, there is the possibility of diagnosing a good posterior approximation for all posterior margins, but not for the joint distribution of model parameters within any model. This could be resolved by constructing a suitable multivariate diagnostic and test.

{\tt R} code for implemeting Algorithm \ref{alg:mc} can be found in the {\tt abc} package, which is freely available on the CRAN.
\\

{\noindent\it Preprint note: The {\tt R} code is currently being incorporated into the above package. In the meantime, it is directly available from {\tt http://www.maths.lancs.ac.uk/$\sim$prangle/pub.html}}

\section*{Appendix: Proofs}

Here we provide proofs of Results \ref{res:postcov_pi}--\ref{res:equivcov_mc} presented in Section \ref{sec:coverage}.  Denote by $F(m, \theta, y)$ the joint distribution function defined by $\pi(y|\theta,m)\pi(\theta|m)p(m)$. We will also use $F$ to denote the associated marginal and conditional distributions.

\paragraph{Proof of Result \ref{res:postcov_pi}:} We have $g(\theta|y)=\pi(\theta|y)$ and $h(\theta_0|y_0)=\pi(\theta_0|y_0)$. In this case, $G_y(\theta) = F(\theta | y) = H(\theta | y)$.  Hence
\begin{eqnarray*}
 & & \Pr(\theta_0 \in C(y_0, \alpha) | y_0) = 
\Pr(G_{y_0}(\theta_0) \in B(\alpha) | y_0) = \alpha \\
&\Rightarrow & \Pr(\theta_0 \in C(y_0, \alpha)) = \E_{G(y_0)}[\Pr(\theta_0 \in C(y_0, \alpha) | y_0)] = \alpha.
\end{eqnarray*}

\paragraph{Proof of Result \ref{res:priorcov_pi}:}
We have $g(\theta|y)=\pi(\theta)$ and $h(\theta_0)=\pi(\theta_0)$. In this case, $G_y(\theta) = F(\theta) = H(\theta)$.  Hence
\[
\Pr(\theta_0 \in C(y_0, \alpha)) =
\Pr(G_{y_0}(\theta_0) \in B(\alpha))
= \alpha.
\]

\paragraph{Proof of Result \ref{res:equivcov_pi}:}

First assume that coverage holds.  Let $B(\alpha) = [0,\alpha)$. Then
\[
\alpha = \Pr(\theta_0 \in C(y_0, \alpha)) = \Pr(p_0 \in [0, \alpha)), \qquad \text{(applying $G_{y_0}$ to the event)}
\]
so the distribution function of $p_0$ equals that of a $U(0,1)$ distribution.
For the converse, now assume that $p_0 \sim U(0,1)$.  Then
\begin{align*}
\alpha = \Pr(p_0 \in B(\alpha)) = \Pr(\theta_0 \in C(y_0, \alpha)), \qquad \text{(applying $G_{y_0}^{-1}$ to the event)}
\end{align*}
which is the condition needed for coverage.

\paragraph{Proof of Result \ref{res:postcov_mc}:}

We have $g(m|y) = p(m | y)$ and $h(m_0|y_0)=p(m_0|y_0)$. In this case, $g(m|y) = p(m | y) = h(m | y)$.  Fix some $i$ and $I$ such that \eqref{eq:Apos} does not hold, and write $A$ for $A(i,I)$.  Then for $y_0 \in A$, $h(i|y_0) = g(i| y_0) \in I$.  Thus $\Pr(m_0=i | y_0 \in A) = \E_{H(y_0)}[h(i | y_0)] \in I$, where $H(y_0)$ denotes the marginal distribution of $H$. Hence \eqref{eq:mc_cov} holds.

\paragraph{Proof of Result \ref{res:priorcov_mc}:}

We have $g(m|y) = p(m)$ and $h(m_0)=p(m_0)$. In this case, $g(m| y) = p(m) = h(m)$.  Fix some $i$ and $I$ and write $A$ for $A(i,I)$.  Consider first that $p(i) \notin I$.  Then $A = \emptyset$ and so \eqref{eq:Apos} holds.  Suppose instead that $p(i) \in I$.  Then $A$ is the set of all possible $y$ values, and $\Pr(m_0=i | y_0 \in A) = \Pr(m_0=i) = p(i) \in I$. Hence \eqref{eq:mc_cov} holds.

\paragraph{Proof of Result \ref{res:equivcov_mc}:}

Assume that, for all $i$, (\ref{eqn:ptest2}) holds for almost all $w$ (with respect to $\mathcal{Z}^{(i)}$ as defined in the statement of this result).
 Fix some $i$ and $I$, and consider the case where \eqref{eq:Apos} does not hold.  Then
\begin{eqnarray*}
\Pr(m_0=i | y_0 \in A(i,I)) &=& \Pr(m_0=i | z_0 \in I) \\
&=& \E_{Z'(z_0)}[p(i|z_0)] \\
&=& \E_{Z'(z_0)}[ z_0 ] \in I,
\end{eqnarray*}
where $Z'(z_0)$ is the marginal distribution of $z_0$ truncated to $I$.  Thus, coverage holds.

Next assume coverage with respect to $H$, and fix some $i$.
For any $w$ such that $\Pr(z_0 = w) > 0$, it is immediate that \eqref{eqn:ptest2} follows.  Define $I_\varepsilon(w) = [w-\varepsilon, w+\varepsilon] \cap [0,1]$. 
It suffices to prove that \eqref{eqn:ptest2} holds for $w$ such that $\Pr(z_0 \in I_\varepsilon(w)) > 0$ for all $\varepsilon > 0$. (The set of other $w$ values has probability zero, as each $w$ lies within an interval of zero probability.)  Fix such a $w$ with $\Pr(z_0 = w) = 0$.  Note that $z_0 \in I_\varepsilon(w)$ represents the same event as $y_0 \in A_\varepsilon := A(i, I_\varepsilon(w))$.  From the assumption on $w$, \eqref{eq:Apos} is false for $I=I_\varepsilon(w)$ and $\varepsilon>0$.  Hence \eqref{eq:mc_cov} must hold i.e.
\[
\Pr(m_0=i | y_0 \in A_\varepsilon) \in I_\varepsilon(w) \qquad \text{for } \varepsilon>0.
\]
The left hand side of this equals
\[
\Pr(m_0=i | z_0 \in I_\varepsilon(w)) = \E_{Z'_\varepsilon(z_0)}[\Pr(m_0=i | z_0)],
\]
where $Z'_\varepsilon(z_0)$ is the marginal distribution of $z_0$ truncated to $I_\varepsilon(w)$.  
Thus
\[
\E_{Z'_\varepsilon(z_0)}[\Pr(m_0=i | z_0)] \in I_\varepsilon(w).
\]
It follows by the Lebesgue differentiation theorem, using the assumption on $z_0$ assumed in the statement of Result \ref{res:equivcov_mc}, that $\Pr(m_0=i | z_0=w) = w$ for almost all $w$, as required.


\bibliography{coverage}


\begin{sidewaysfigure}[p] \begin{center}
\includegraphics{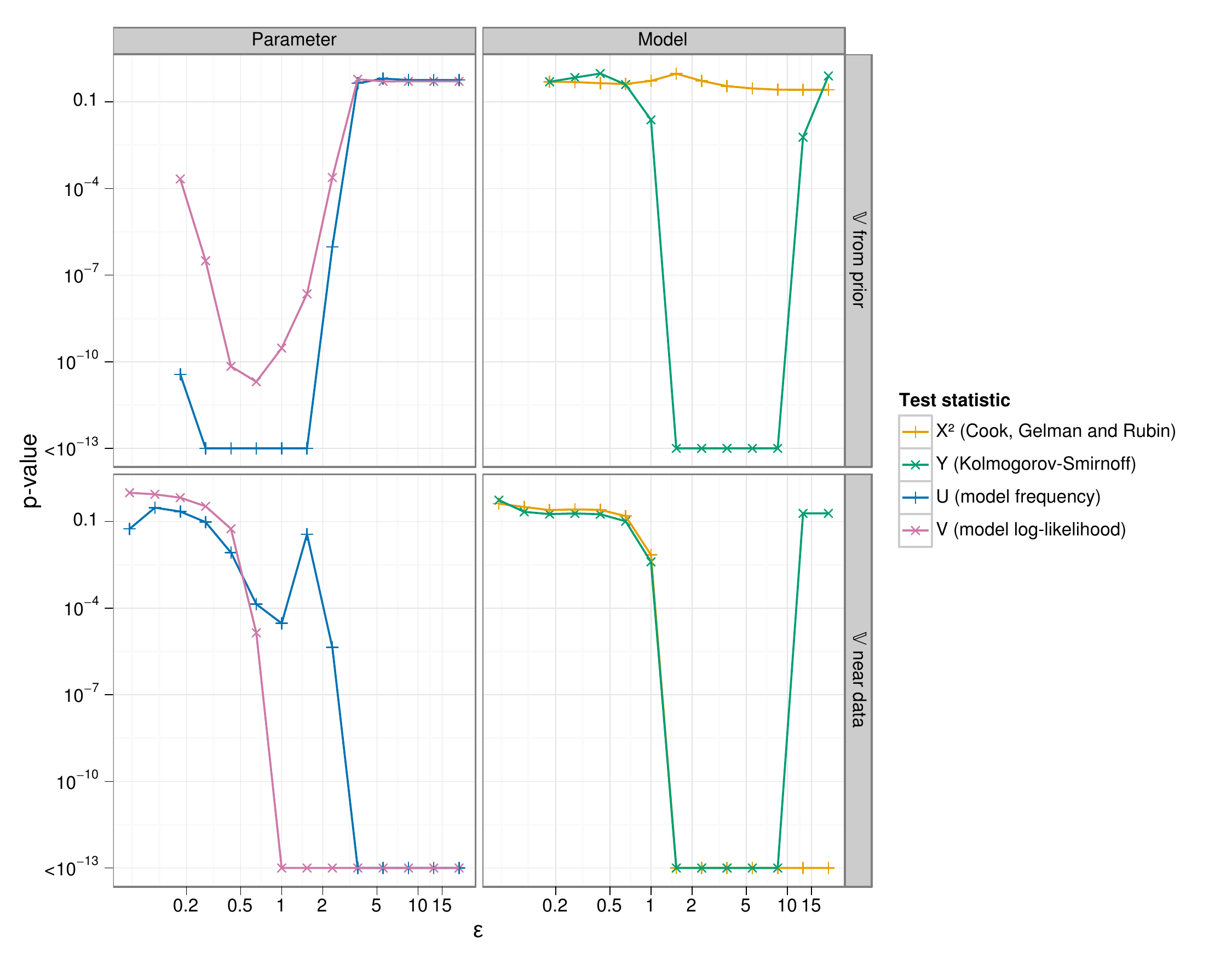}
  \footnotesize{\caption{(Supplementary Figure 1): As Figure 1 (main text), but with $c=500$.
  }} \label{S1}

\end{center}
\end{sidewaysfigure}

\begin{sidewaysfigure}[p] \begin{center}
\includegraphics{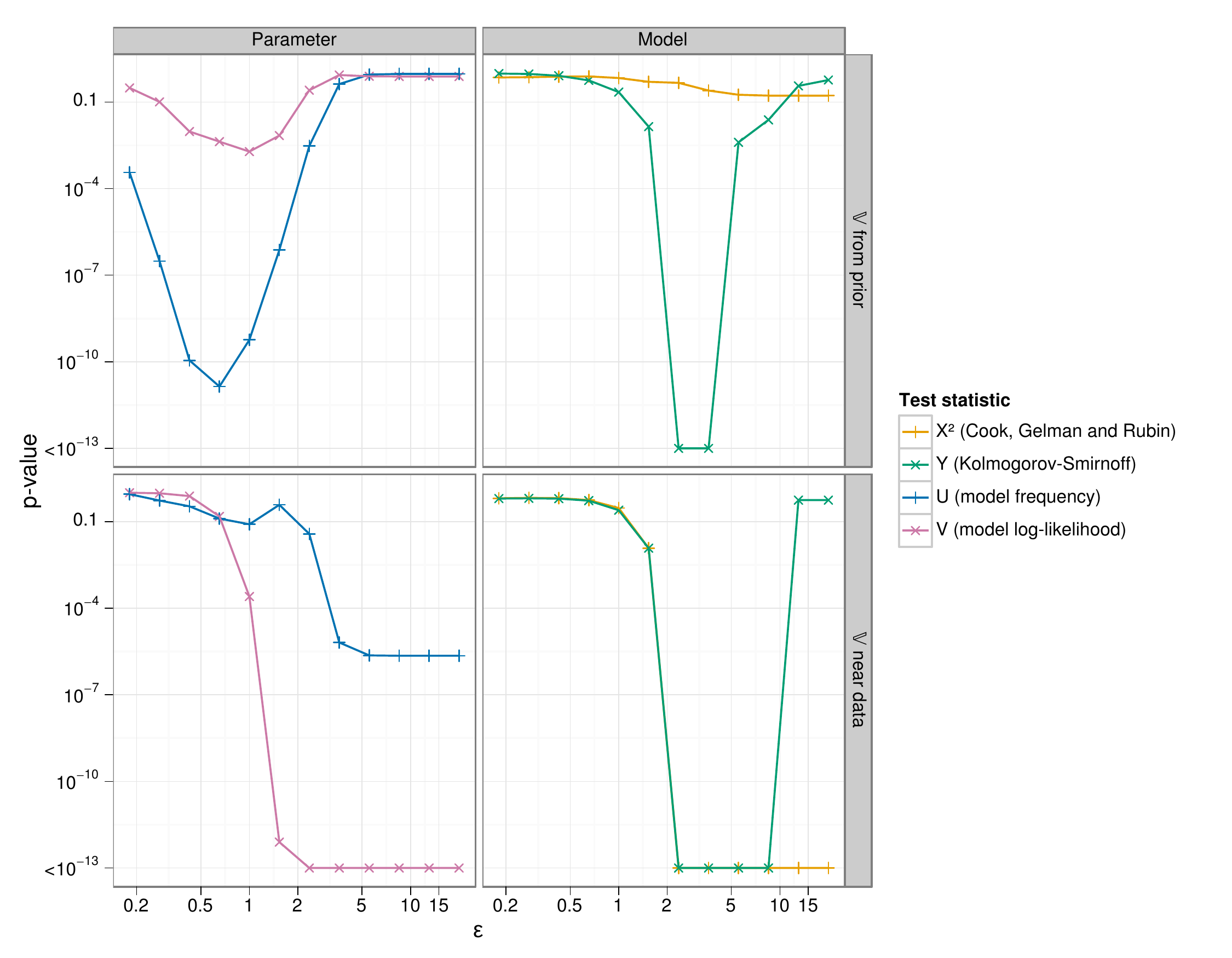}
  \footnotesize{\caption{(Supplementary Figure 2): As Figure 1 (main text), but with $c=100$.
  }} \label{S2}
\end{center}
\end{sidewaysfigure}

\begin{figure}[htp] \begin{center}
  \includegraphics{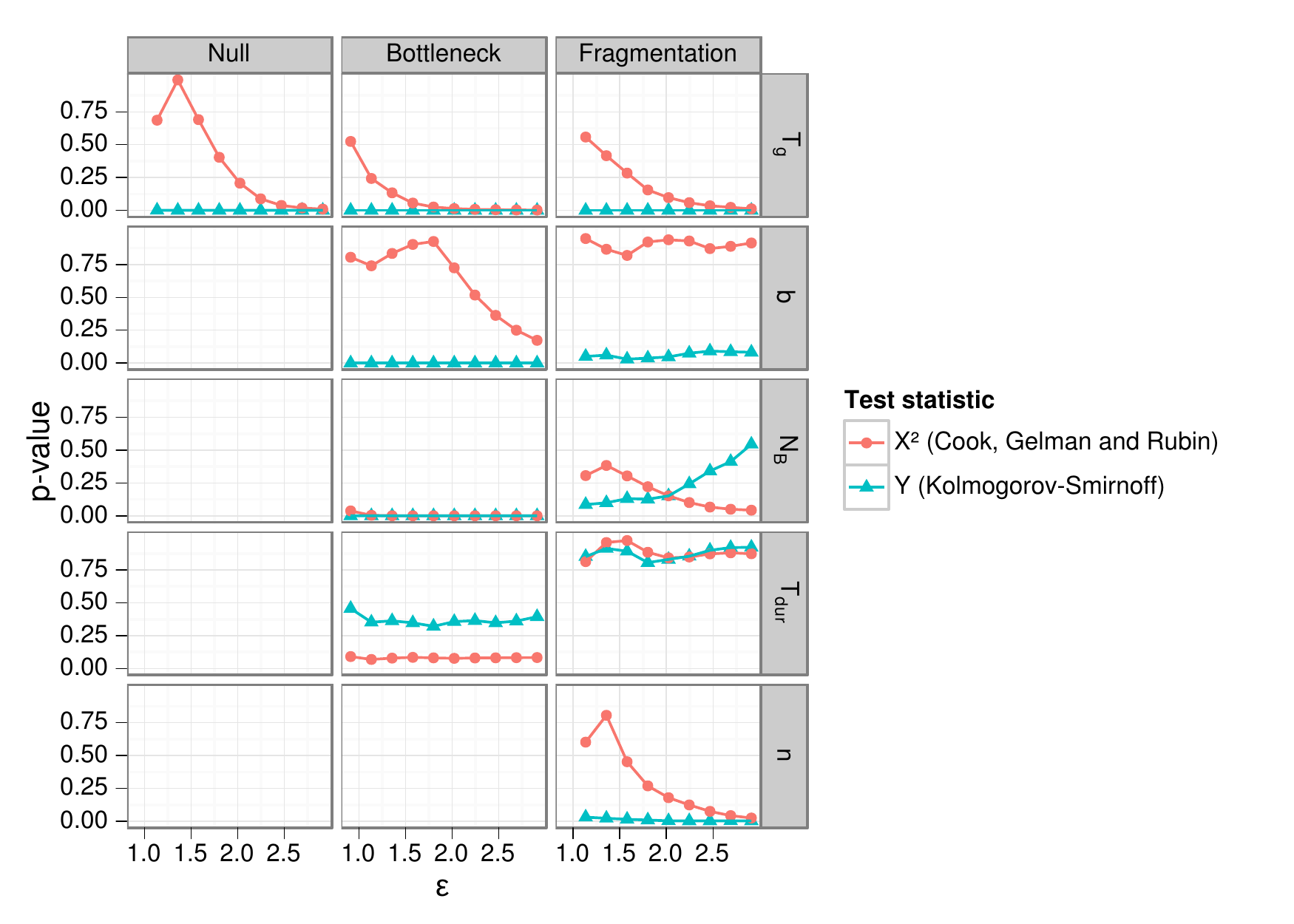}
  \footnotesize{\caption{(Supplementary Figure 3): As for Figure 4 (main text), but for the remaining model parameters $T_g$, $b$, $N_B$, $T_{\text{dur}}$ and $n$.
  } \label{S3}}
\end{center}
\end{figure}

\begin{figure}[htp] \begin{center}
  \includegraphics{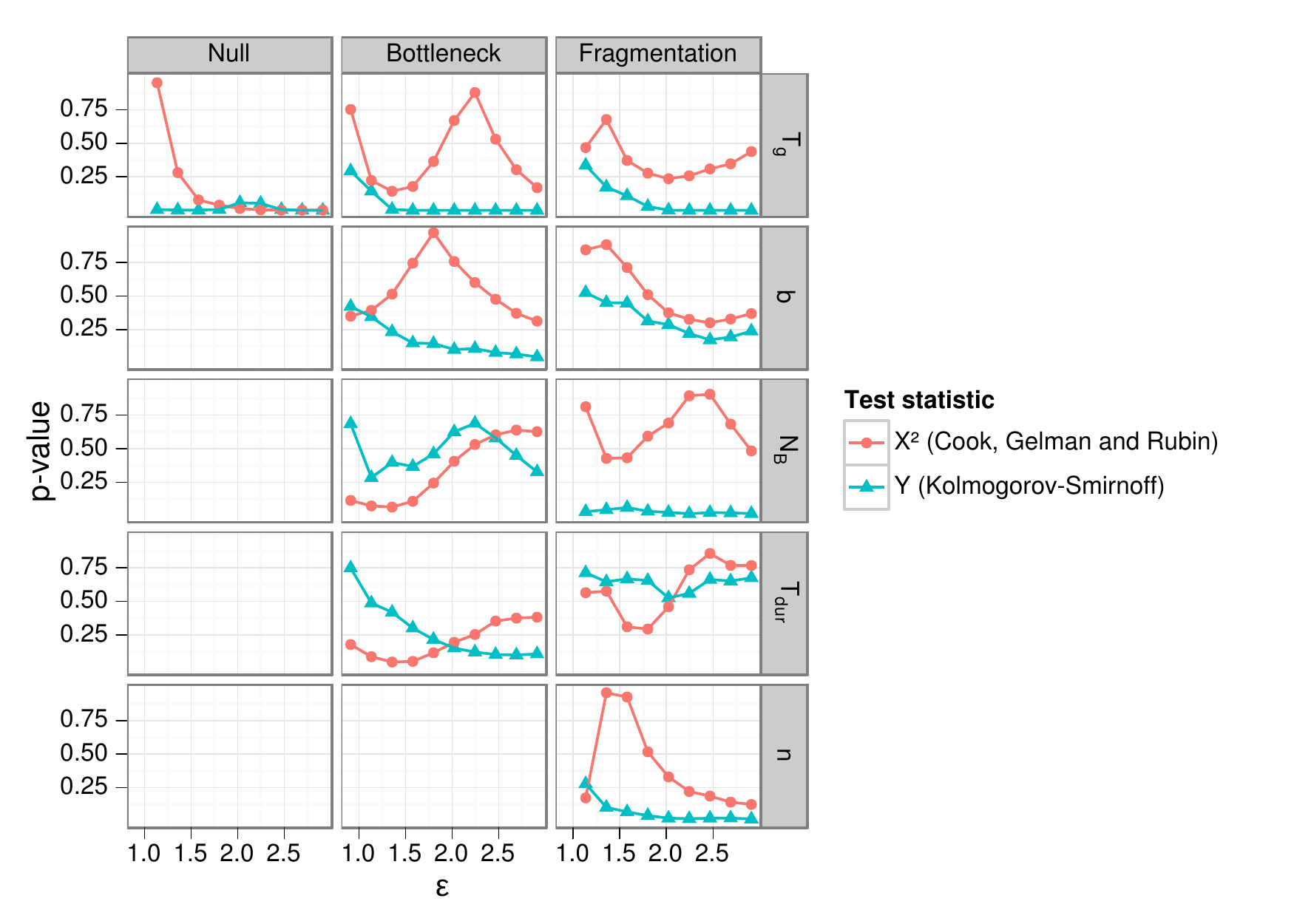}
  \footnotesize{\caption{(Supplementary Figure 4): As for Figure 5 (main text), but for the remaining model parameters $T_g$, $b$, $N_B$, $T_{\text{dur}}$ and $n$.
  }} \label{S4}
\end{center}
\end{figure}

\begin{figure}[!p] \begin{center}
  \includegraphics{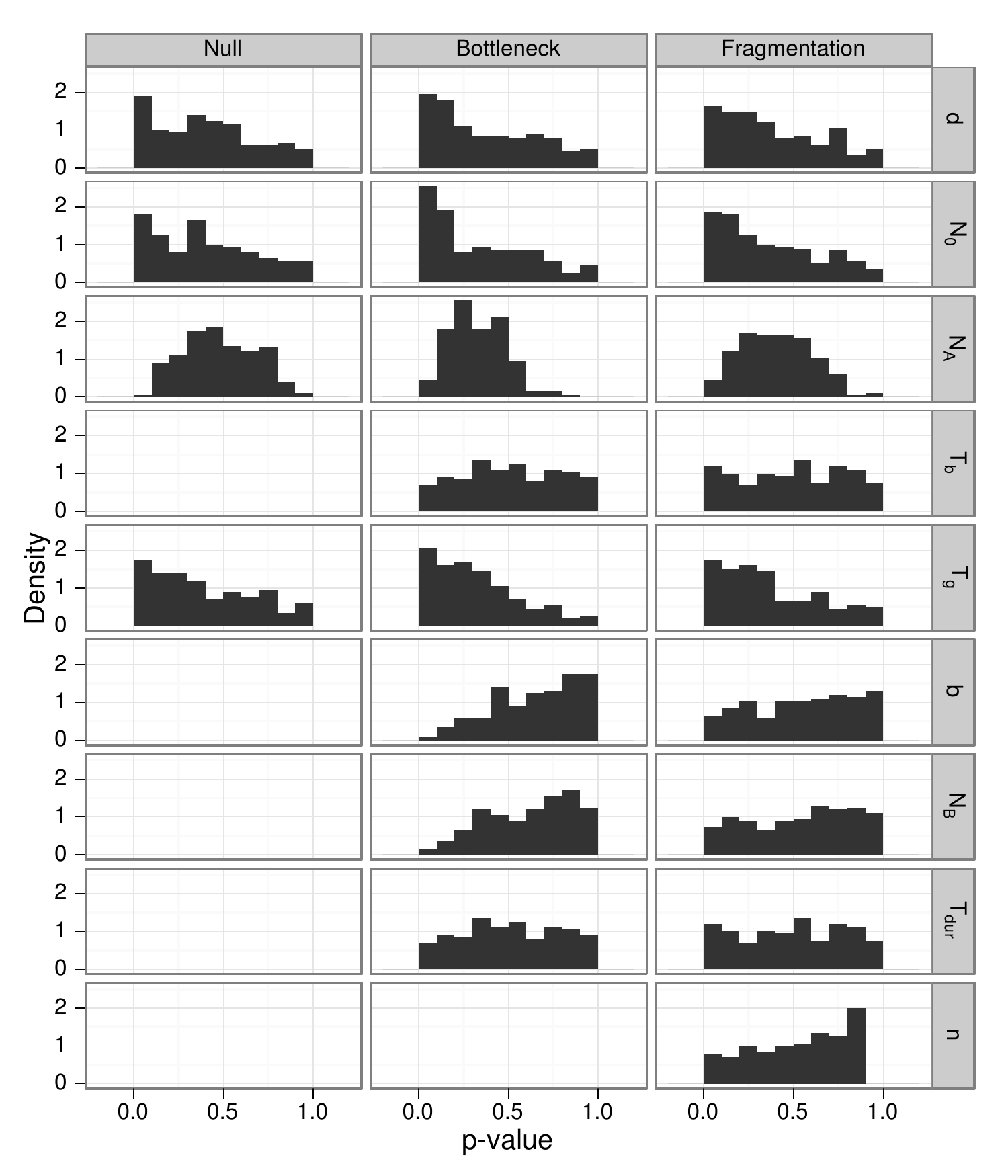}
  \caption{\footnotesize{(Supplementary Figure 5): Histograms of the $c=200$ $p_0$ values for the parameters $d$, $N_0$, $N_A$ and $T_b$ in the human demographic history analysis, with $\epsilon = 1.36$. Regression-adjusted post-processing is not implemented. Rows correspond to individual parameters; columns correspond to the three models.
   }}
  \label{S5}
\end{center}
\end{figure}

\begin{figure}[!p] \begin{center}
  \includegraphics{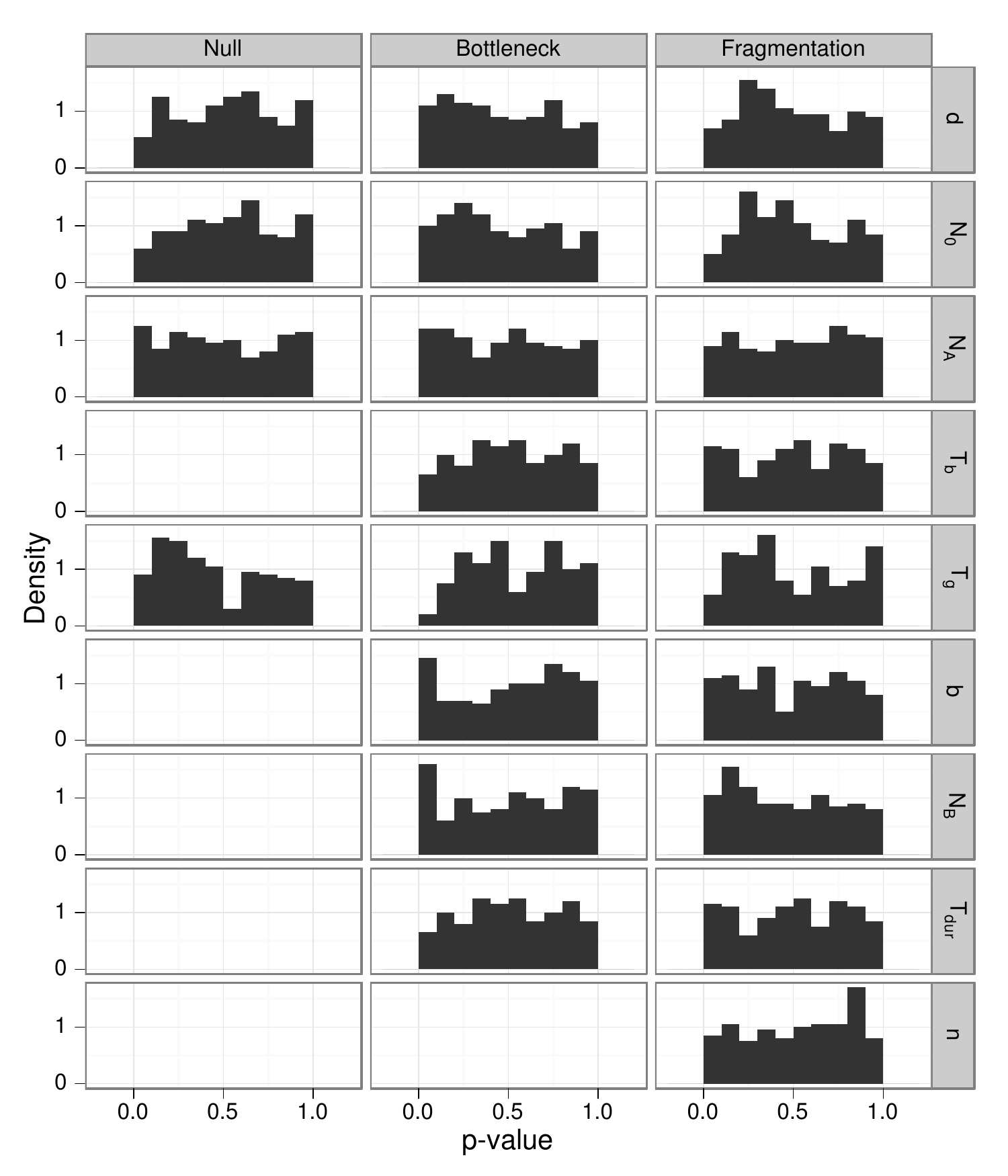}
  \caption{\footnotesize{(Supplementary Figure 6): Histograms of the $c=200$ $p_0$ values for the parameters $d$, $N_0$, $N_A$ and $T_b$ in the human demographic history analysis, with $\epsilon = 1.36$. Regression-adjusted post-processing has been implemented. Rows correspond to individual parameters; columns correspond to the three models.
   }}
  \label{S6}
\end{center}
\end{figure}

\end{document}